# Urban Flood Drifters (UFDs): identification, classification and characterisation

Arnau Bayón[1], Daniel Valero[2,3,*], Mário J. Franca[2]

[1] Department of Hydraulic Engineering and Environment, Universitat Politècnica de València (Spain)

[2] Institute for Water and Environment, Karlsruhe Institute of Technology (Germany)

[3] Currently: Department of Civil and Environmental Engineering, Imperial College London (United Kingdom)

* Corresponding author: Daniel Valero (daniel.valero@kit.edu, d.valero@imperial.ac.uk )

**Abstract**

Extreme floods threaten lives, assets and ecosystems, with the largest impacts occurring in urbanised areas. However, flood mitigation schemes generally neglect the fact that urban floods carry a considerable amount of solid load. In this study, we define Urban Flood Drifters (UFDs) as loose objects present in the urban landscape that can become mobile under certain flow conditions, thereafter blocking drainage infrastructure and endangering both downstream and upstream communities. Based on 270 post-flood photographic records from 63 major inundations of the past quarter-century across 46 countries, we provide a comprehensive analysis of UFDs and their flood-hazard implications. We show that a variety of vehicles, furniture and a heterogeneous mixture of drifters are present in post-flooding scenarios. Plastic, construction debris and wood (natural or anthropogenic) dominate the statistics of transported drifters in urban floods (with frequencies of roughly 50—60% each), followed by cars (present in 31.5% of post-flood images). Other heavy vehicles are readily observed in post-flood imagery and furniture such as bins, garden sheds or water tanks also appear occasionally, therefore suggesting that they can play a relevant role in extreme floods.

1. **Introduction**

Floods are among the top three most common and severe weather-caused events (World Economic Forum, 2022) and will be responsible for the displacement of over 200 million people by 2050 (Clement et al., 2021). Furthermore, climate change, human activities and land-cover alteration are making floods more intense, frequent, and unpredictable (Ludwig et al., 2023; United Nations, 2020a). The economic consequences of floods in recent decades have been staggering. According to the European Union (European Environment Agency, 2022), hydrometeorological events account for roughly two-thirds of overall damage caused by natural disasters since the 1980s. The European Union reports that a single flood event can result in direct losses of billions of euros (European Environment Agency, 2022), not to mention the devastating impact on human life, the environment, and cultural heritage (Jonkman and Vrijling, 2008; Hickey and Salas, 1995; Molinari et al., 2020; Arrighi, 2021; Bates et al., 2023). In low- and mid-income countries, floods are responsible for more than half of weather-related disasters (United Nations, 2020b). These most vulnerable regions also have fewer resources to plan, build, maintain, and upgrade flood management infrastructure, which can create a challenge at all levels of governance (Zevenbergen et al., 2008, 2010; Fang, 2016; Smith et al., 2019a; Wing et al., 2022).

The acute consequences of flooding during the past few decades have prompted numerous investigations on urban flooding (Mignot et al., 2019; Smith et al., 2019a; O'Donnell and Thorne, 2020; Zevenbergen et al., 2020; Wing et al., 2022; Bates et al., 2023; Sanders et al., 2023). However, it was not until recent floods that significant attention was drawn to the role of drifters carried by the floods (Dewals et al., 2021; Bung, 2021; Xiaotao and Jindong, 2021; Dick et al., 2021; Dietze and Ozturk, 2021; Apel et al., 2022; Mohr et al., 2023; Ludwig et al., 2023). Recent studies have emphasized the importance of considering drifters conveyed by large floods when evaluating urban flood hazards (Dewals et al., 2021; Mohr et al., 2023). Flood consequences can be more intense when large loose objects, such as vehicles or urban/household furniture, are washed away. However, these are usually ignored in flood analyses and a substantial part of the scientific literature (Smith et al., 2019a; Bates, 2022; Pregnolato et al., 2022; Sanders et al., 2023) and, therefore, their role remains notably unexplored. Literature may have neglected this phenomenon, either wholly or partially, since understanding its dynamics is exceptionally intricate and heavily reliant on case-specific scenarios. Assessing the impacts of drifters during floods depends on various factors, including flood conditions, drifter characteristics, and fluid-solid interactions.

We define Urban Flood Drifters (UFDs) as objects present in the urban landscape, (i) which are not moored or fixed and that remain stable under dry conditions; (ii) but which may become unstable when exposed to high water levels and/or flow velocities. For some UFDs, even a few centimetres of water depth can lead to their destabilisation and motion (Keller and Mitsch, 1992; Kramer et al., 2015; Martínez-Gomariz et al., 2018; Smith et al., 2019b; Martínez-Gomariz et al., 2020; Shah et al., 2021), thus becoming part of the flood and contributing to the worsening of its consequences.

At high velocities, UFDs have the potential to cause significant injuries to pedestrians, and damage to private property and infrastructure (Jalayer et al., 2018; Zhang et al., 2018, Xiong et al., 2022). However, due to the limited investigation of UFDs in the literature, the full extent of their role in flooding and their contribution to flood hazards is not entirely clear. Drifters-induced blockages can hinder flow at bridges, tunnels, culverts, and other singularities and cause constrictions of the hydraulic drainage infrastructure (Figure 1); in these, obstruction occurs through the formation of a heterogeneous jam of drifters, that acts as a porous dike, thereby increasing upstream water levels due to the backwater effect (Figure 2). UFDs mobilisation hence increases the inundation hazard for homes and businesses, disrupts pedestrian and vehicle traffic, and hinders emergency response teams. Infrastructure obstructions, if they overtop and/or collapse, can eventually induce dam break-type waves downstream. Urban drifters may also carry pollutants such as oil, chemicals, plastic, metals, and organic waste. These pollutants may be further washed into the environment, causing harm to aquatic habitats and wildlife, and compromising the water quality.

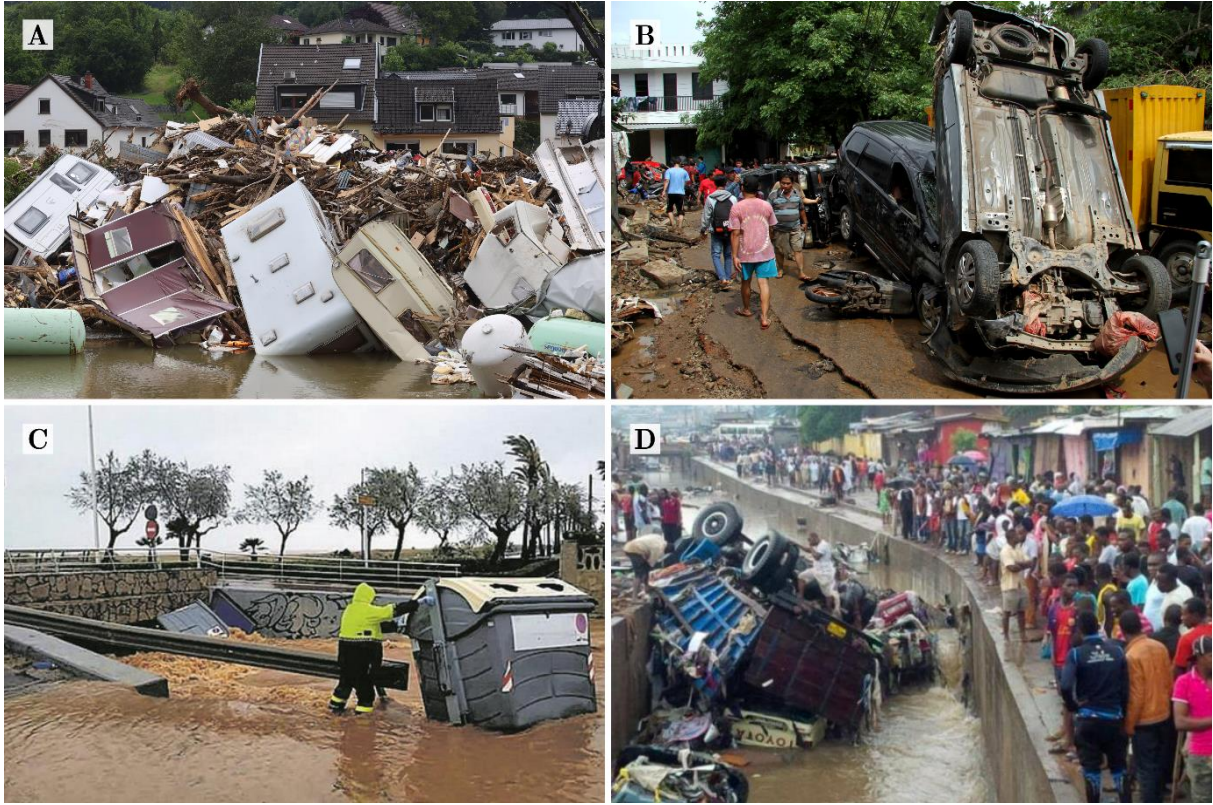

*Figure 1.* Disruptions during major floods caused by urban flood drifters forming dikes and clogging urban drainage systems: A Caravans, fluid tanks, garden sheds and wood drifters in the July 2021 floods in the Ahr river, Germany (Rattay, 2021), B Cars and truck trailers in Jakarta, Indonesia (Andrianto, 2020), C Dumpsters in Benicarló, Spain (Sánchez Albiol, 2020) and D Cars, trucks and small drifters in Accra, Ghana (Modern Ghana, 2018). Images reproduced with permission; copyright corresponds to the endorsed sources.

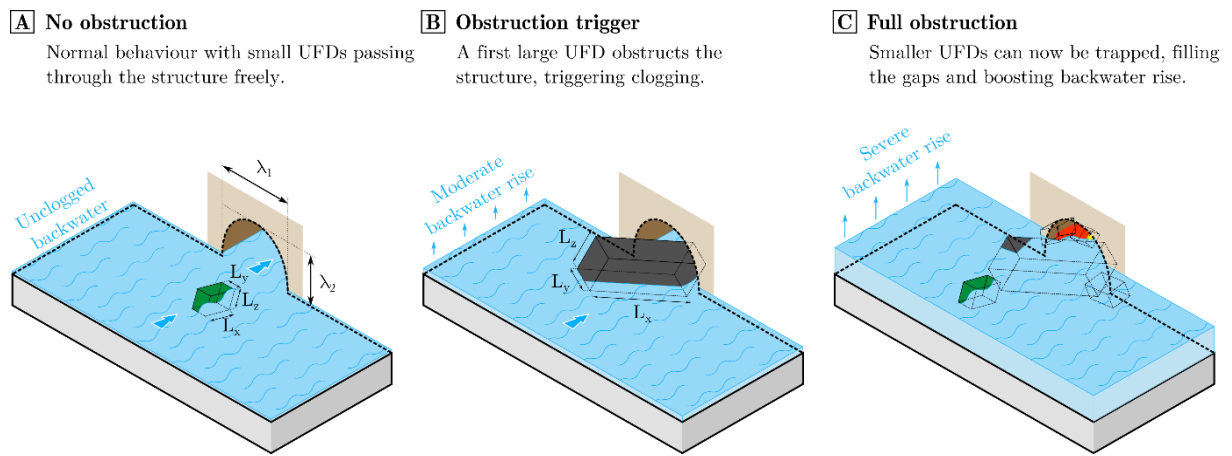

*Figure 2.* UFDs-jam formation during the blockage in a flow contraction of the channel network and subsequent backwater effect. $\lambda_1$ and $\lambda_2$ are the main contraction dimensions, and $L_x$, $L_y$ and $L_z$ the main drifter dimensions.

The drainage capacity can be reduced by transported and deposited UFDs, consequently amplifying flood hazards beyond what was anticipated during planning phases, where flood modelling is often performed only considering the water phase over fixed geometries or with little information about what can potentially block drainage infrastructure. This *only-water* hypothesis is simplistic or unsubstantial since vehicles, urban/household furniture, common in any urban landscape, are also subject to becoming a moving part of the flood. Figure 1 shows post-event images of major floods around the world, where the ubiquitous presence of very diverse UFDs is observed. In particular, it is observed that the presence of UFDs in flood events has devastating consequences in different urban environments, regardless of geography, demographics, or climate. Large urban objects, that can potentially become UFDs, are heterogeneous in nature and can vary in shape, size, and weight. An increasing number of studies in the literature are tackling the analysis of debris mobilization by floods through various approaches. Some authors employ advanced numerical methods to investigate this phenomenon (Xiong et al., 2022; Hasanpour et al., 2023). However, a proper topology and characterisation of these drifters does not exist yet, contributing to the lack of knowledge of their influence and feedback on the hydrodynamics of flood propagation, and resulting consequences in urban flood hazard.

With this investigation, a systematic analysis of UFDs is carried out, aiming at their characterisation and classification, and ultimately at the quantification of their prevalence in urban floods and stability characteristics. The results, herein presented and discussed, can contribute to the potential development of reliable well-informed modelling tools for flood managers. This study has three specific objectives:

- *Identify* urban objects that are susceptible to becoming UFDs.
- *Classify* different types of UFDs to facilitate future systematic studies.
- *Quantify* the prevalence of UFDs after major floods and their flood-risk characteristics.

## 2. Identification and classification of UFDs

**Data acquisition protocol**

After a large flood event, numerous media reports are published with images that contain valuable information (Valero et al., 2021). These images can be captured by professional photographers or passers-by, and often remain hosted online on news platforms or social media. They facilitate a prompt depiction of the consequences on-site, although they may hold a large bias towards the most dramatic aspects of the disaster. In order to provide relevant information from that available imagery, we define a protocol that focuses on the identification of drifters transported by floods and that allows the quantification of their prevalence in the post-flood urban landscape. With this information, we can later define an informed classification that covers the wide spectrum of UFDs and characterise each group in terms of their main flood-transport properties, based on available product catalogues.

The image-collection and drifter-detection protocol is presented below, step-by-step:

1. With the aid of Google's (.com) search engine and inspired, but not exclusively targeting, a list of "deadliest floods" (Wikipedia Commons, 2022) including countries, years, and even months, and considering the range of years from 1999 to 2023, we search for the following keywords: "flood", a given country, and a given year and month in which the flood occurred; identifying at least 10 images from major floods every year.

2. We prioritise, in the following order: newspapers, magazines, articles, and blogs; and find different sources, storing their images in local folders. All Uniform Resource Locators (URLs) are stored in our electronic supplementary material (Bayón et al., 2024). When building the image-bank, we bear in mind the following considerations: a) images that do not depict visible drifters are not considered, and b) duplicates are avoided, i.e. in the case that photographs show the same scenario and drifters, but from different angles.

3. We repeat the previous steps until we have stored 10 images for each year, at least. When not possible, to prevent a lack of representation of years with few significant floods or lack of graphic material, we also use the search engine of YouTube, following the criteria presented in step (i), and inspect videos frame-by-frame to provide at least one photographic record where drifters are visible.

The data used in this study ($N_{img}$ = 270 retrieved images) were extracted from online available imagery and represent 63 major floods in 46 different countries over 25 years. A simplified analysis with a reduced number of images ($N_{img}$ = 200), showed little differences in the prevalence of the classes of UFDs. The geographic distribution outlined in Table 1 reveals that regions with higher income levels receive more extensive media coverage after flood events. For instance, over half of the pictures were taken in Europe and Northern America. Additionally, demographic density is a significant factor, as more densely populated areas tend to receive wider coverage. For example, Asia and the Pacific, which represents approximately 57% of the world's population (World Bank, 2021), accounts for 31.1% of the images. Looking separately by continent reveals that no big differences are present in the prevalence of classes, whereas groups may differ; for instance, vehicles' frequency in floods are similar, but caravans & RVs are more common in high-income regions whereas two-wheelers appear more often in low- and mid-income areas. The data are available online (Bayón et al., 2024), including URLs to the original sources and images.

| **Region** | **Frequency** |
|---|---|
| Western European and Others | 51.9% |
| Asia and the Pacific | 31.1% |
| Latin America and the Caribbean | 7.8% |
| Africa | 5.2% |
| Eastern Europe | 4.1% |

*Table 1. Frequency of analysed post-flood imagery during the 1999–2023 period, using regional groups defined by United Nations (2023).*

**Limitations**

While our method offers valuable insights into the presence of drifters or litter in urban environments (i.e., UFDs) following flood events, it has some limitations. Firstly, the method focuses on post-flood pictures exclusively featuring UFDs, and therefore neglects images without UFDs, which means that our study discloses how often a certain UFD is transported by a large flood, but not how much of it is mobilised or their associated flood hazard. Likewise, the reliance on press and social media platforms as data sources introduces an additional bias, as these platforms tend to highlight on only a portion of reality and may not depict the complete picture of post-flood scenarios.

The choice of English as the search language is another constraint, despite its global prevalence, since local media are first respondent to floods. This linguistic bias may result in overlooking relevant data in other languages, potentially limiting the comprehensiveness of our analysis.

Additionally, a notable over-representation of high-income countries in our dataset introduces a geographical bias, with the identified UFDs being an impression of local socio-cultural circumstances (c.f., types of vehicles or level of fixture of urban furniture). As disasters are more extensively covered in these regions, our findings may not fully capture the scenario in mid- and low-income countries, leading to their potential underrepresentation. Future research efforts are needed to address these limitations.

**Data clustering**

Overall, the data mining process resulted in a total of 270 images, which were organized according to year and country of occurrence. Each image contained one or more groups of UFDs that were visually identified and flagged in tables according to very specific terms. In that process, even if an image has a large amount of plastic (for instance), this belongs to one group of UFDs (later introduced in Table 2) and only one is counted, since the goal is to disclose how often different types of UFDs are affected by floods, but not to estimate the total amount transported by a given flood (not identifiable in images). Within the 270 images retrieved, a total of 756 UFDs where identified (i.e., one group in an image is counted as 1, regardless of the amount of them in that image). Afterwards, a more formal categorization was defined, clustering UFDs based on their size, function and abundance in the urban landscape. This final classification includes three main classes of UFDs:

- *Vehicles*, of all types with or without engine, and any number of wheels. This can include from e-scooters and bikes, to caravans, buses, trucks and other heavy vehicles.
- *Furniture*, of private or public origin.
- *Heterogeneous drifters*, often of unidentifiable origin.

The first two classes generally comprise a family of generally large-sized items, that have dimensions comparable to the contraction dimensions of culverts and bridges ($\lambda_1$, $\lambda_2$ in Figure 2). These objects have typical standardized (typified) physical characteristics and are of industrial origin, and so their mechanical characteristics, such as dimensions or weight, are catalogued (e.g., vehicles or furniture). The third class includes drifters of smaller size and more heterogeneous nature, of industrial or natural origin, which can also obstruct certain intake structures; for example, Honingh et al. (2020) confirmed that small plastic particles can significantly clog urban drainage grates.

Each of the three proposed UFD classes can be further divided into more specific groups, based on characteristic physical properties, such as size (e.g., for vehicles: bikes vs. trucks), function (urban vs. household furniture) or material (e.g., wood vs. plastic). Table 2 provides the herein proposed classification and grouping of UFDs, and Figure 3 illustrates the subdivision in classes and groups of the typified UFDs.

| Class | ID | Group | Description |
|---|---|---|---|
| Typified UFDs — Vehicles | V1 | Two-wheelers | Bikes, motorbikes and e-scooters. |
| | V2 | Cars | Cars and other light four-wheel vehicles designed to transport passengers. |
| | V3 | Vans | Vans and other heavy four-wheel vehicles designed to transport materials and stock. |
| | V4 | Caravans & RVs | Vehicles designed to provide habitable space. |
| | V5 | Large heavy vehicles | Vehicles designed to transport many people or goods (buses, trucks, trains, boats, etc.). |
| Typified UFDs — Furniture | F1 | Urban furniture | Facilities designed to provide a public service in streets (bins, traffic barriers, etc.). |
| | F2 | Household furniture | Facilities from private front gardens that can be dragged by floods (tanks, garden sheds, etc.). |
| Heterog. UFDs | DC | Construction | Debris that can be dragged from construction sites or damaged buildings. |
| | DM | Metal | Metal debris, predominantly of constructive origin (sheets, pipes, etc.). |
| | DP | Plastic | Plastics and textile objects of small dimensions and irregular shape. |
| | DW | Wood | Natural wood (trunks, branches, etc.) and processed wood. |
| | DO | Others | Other drifters of uncertain origin (food, tableware, leaves, sediment, etc.). |

*Table 2. Classification and grouping of UFDs based on available post-flood imagery.*

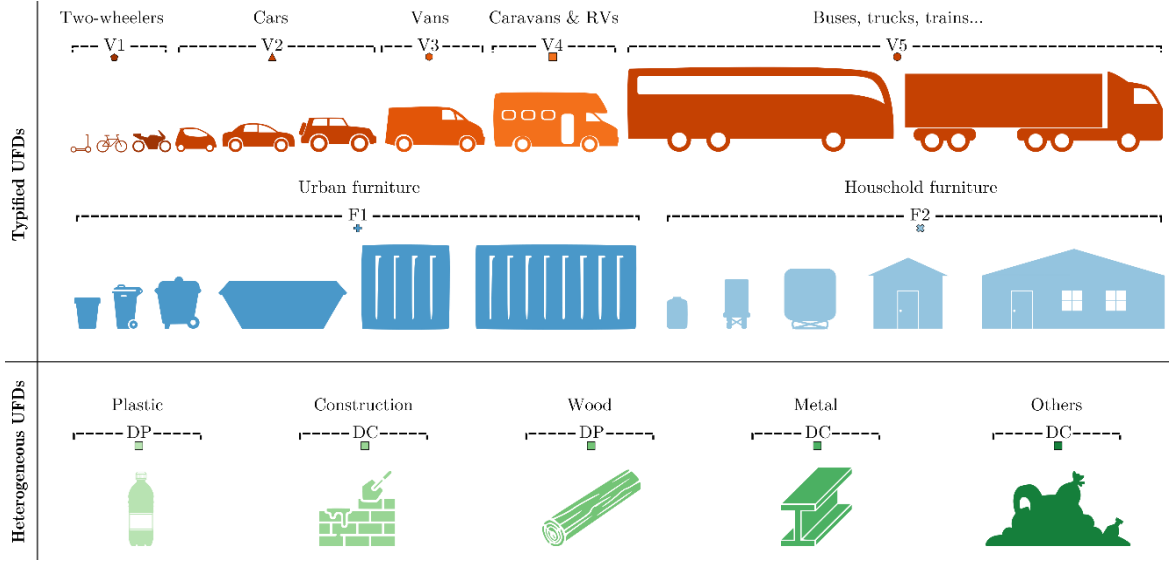

*Figure 3. Typified UFDs: relative sizing and grouping according to the classification presented in Table 2.*

## 3. Characterization of UFDs

**Frequency and prevalence of UFDs in post-flood imagery**

To quantify the presence of UFDs in the analysed post-flood imagery, we define two quantities: frequency (= $N_{UFD,i}/N_{img}$, with $N_{UFD,i}$ the number of images showcasing a UFD of a given group $i$, according to those in Table 2) and prevalence (= $N_{UFD,i}/\sum N_{UFD,i}$). Vehicles are often present in the imagery investigated. In particular, cars (V2) appear in the images with a frequency of 31.5% (Figure 4), whereas other heavier vehicles are also often transported in floods, such as vans (V3, frequency of 5.2%), and trucks and buses (V5, frequency of 5.9%). Larger vehicles such as vans (V3), caravans & RVs (V4) or heavy vehicles (V5), are potentially more hazardous due to their dimensions or larger likelihood to be parked close to rivers, in scenic locations prone to flooding. These are more prone to block bridges, although their stability has not been fully investigated yet (Kramer et al., 2015, 2016). Previous studies on urban flood drifters have almost exclusively focused on cars (Keller and Mitsch, 1992; Al-Qadami et al., 2022; Martínez-Gomariz et al., 2018; Shah et al., 2021; Milanesi and Pilotti, 2020; Shu et al., 2011; Smith et al., 2019b; Xia et al., 2014), which indicates that a portion of other drifters prone to flood-hazard is being overlooked, since 68.5% of images showed other UFDs except cars.

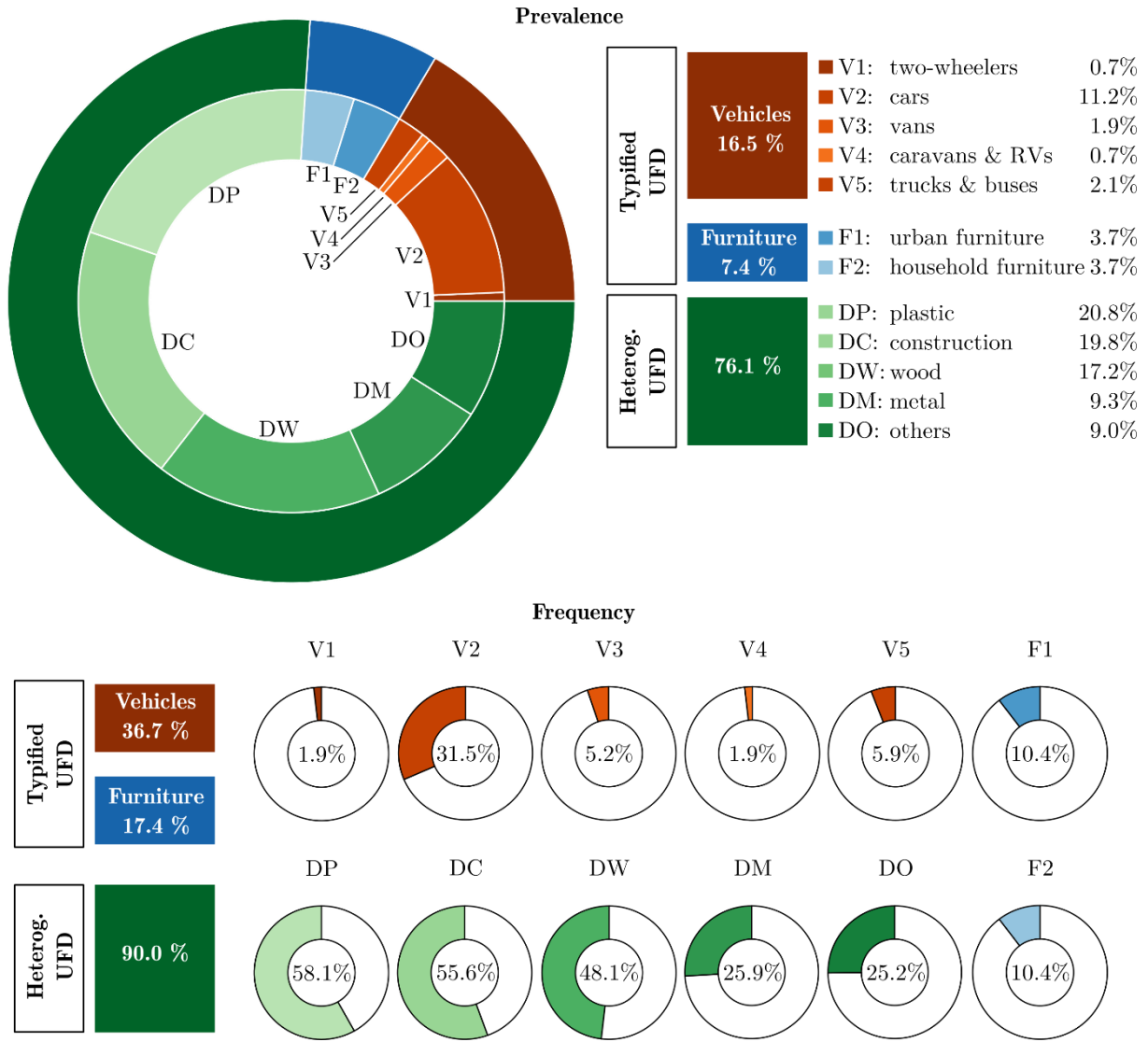

*Figure 4. Prevalence and frequency of UFDs in available post-flood imagery, following UFDs classification and grouping presented in Table 2.*

Within the typified UFDs, urban (F1) and household furniture (F2) are 47% as frequent as vehicles, with 17.4% of the images showing either F1 or F2 mobilised by floods. Only recently, Martínez-Gomariz et al. (2020) addressed the stability of certain UFDs of the group F1 (i.e., dumpsters) and Kramer et al. (2015) their potential clogging of linear infrastructure. Our analysis reveals that they should not be ignored in urban hydrology and flood management studies. This may be especially challenging, as the nature of UFDs types F1 and F2, ranging from small furniture and bins to garden sheds (see Figure 3), is notably more variable than vehicles. Hence, the generalization of the analyses on stability and transport processes becomes more complex.

Heterogeneous drifters, often of small size, are by far the most common in the data mined, with a combined frequency of 90%. Considering the larger mobility of small UFDs and the limitations of our analysis, the amount of plastic and small wood, for instance, may be larger than reported. Heterogeneous UFDs (except some large wood) are generally of small size, which means that alone they can rarely clog the bottlenecks of large drainage infrastructure. However, these can be retained by obstructions caused by larger typified UFDs, clogging these and making them porous dams. Studies on hazards induced by plastic (DP), which are present in 58.1% of the analysed images, are relatively scarce, but show large potential for hazard enhancement by clogging of smaller urban drainage (Honingh et al., 2020).

Studies on wood (DW), which is present in 48.1% of the investigated images, are more common. However, there are still gaps in our understanding of their transport characteristics (Ruiz-Villanueva et al., 2014, 2016; Wohl et al., 2019) and related hazards (Schalko et al., 2018, 2020; Furlan et al., 2019, 2021). To the best of our knowledge, no analogous studies exist that are devoted to other smaller heterogeneous UFD types, which are often present in post-flood imagery, such as construction drifters (DC), metal drifters (DM), or other drifters (DO), with frequencies of 55.6%, 25.9%, and 25.2% in our data.

**UFDs and flood destabilisation hazard**

The stability and transport of UFDs are influenced by various physical variables related to the flow, such as velocity ($v$) and depth ($h$), and to the UFD itself (Keller and Mitsch, 1992; Shu et al., 2011; Kramer et al., 2016; Smith et al., 2019b; Milanesi and Pilotti, 2020; Shah et al., 2021; Al-Qadami et al., 2022). We identify the following forces whose balance determines the UFD's stability under flood conditions: weight ($F_g$), buoyancy ($F_b$), and drag ($F_b$):

$$F_g = Mg; \quad F_g = V_b \rho_w g; \quad F_d = \frac{1}{2} \rho_w C_D A v^2 \tag{1}$$

Friction between the UFD and the floor is also a relevant parameter that opposes flow drag, but it can be explained by a combination of $F_g$ and $F_b$. In the above equations, the physical constants are gravity acceleration ($g$), water density ($\rho_w$), and the UFD's drag coefficient ($C_D$), which may lay closely above or below unity, depending on the specific UFD's shape (White, 2016). The total UFD's mass (M) can be expressed as $M = \rho_m L_x L_y L_z$, where $\rho_m$ is a bulk density that includes all voids within the bounding box, which is defined by the $L$-parameters that are the dimensions of the bounding box (in $x$, $y$, and $z$) containing the UFD (Figure 5). The submerged volume ($V_b$) is the part that contributes to the buoyancy of the UFD and can be expressed as a fraction of the submerged bounding box, $\phi_1 \in (0,1]$, such that $V_b = \phi_1 L_x L_y h$ ($\phi_1 = 1$ represents a solid rectangular prism). The $x$ dimension is considered as the maximum horizontal dimension of the UFD, and therefore the largest face exposed to a flood can be expressed as $A = \phi_2 L_x h$, where $\phi_2 \in (0,1]$ is a geometry-dependent parameter ($\phi_2 = 1$ represents a full rectangular face exposed to the flow).

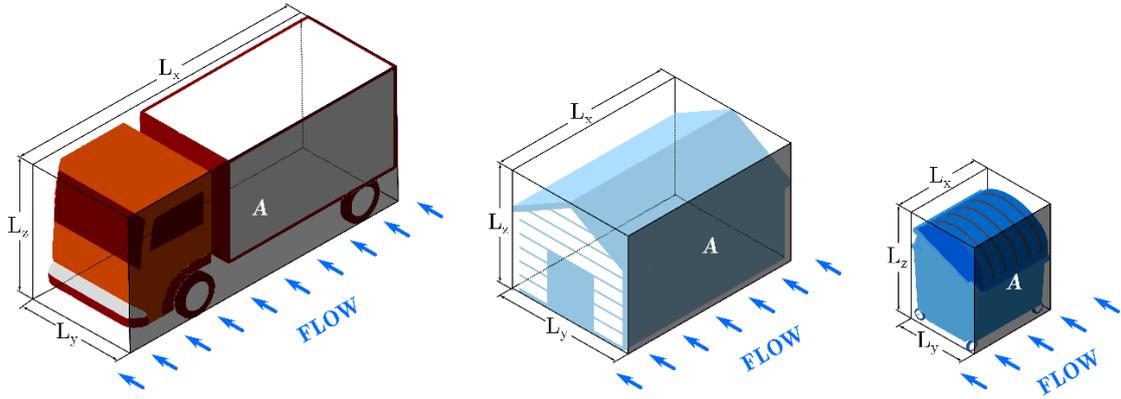

*Figure 5. Examples of UFD dimensions where $L_x$ is the length, $L_y$ is the width and $L_z$ is the height, which form a bounding box with volume $V$, and exposed area to flooding $A$, in the represented case when the UFD is fully submerged ($h = L_z$, and $\phi_2 = 1$). Flow from right to left.*

Based on Equation 1, we quantify the most relevant physical parameters of over 100 potential typified UFDs using available commercial specifications and our own measurements; the latter, by using metric tape and mass scale on objects belonging to categories of Table 2. The parameters under analysis are their length ($L_x$), width ($L_y$), height ($L_z$) and mass ($M$), which allow the calculation of the potential UFD's bounding-box volume ($V$), bulk density ($\rho_m$) and exposed frontal area ($A$, when assuming $\phi_2 = 1$), as depicted in Figure 5. Due to the large variability and potentially easier transport during floods of smaller heterogeneous UFDs, they are kept out of the following analysis.

Figure 6 displays the analysed potential-UFDs based on some key physical properties. We note that increasing $L_x$ has associated an increase in the traverse dimension $L_y$ (Figure 6A), although $L_y$ saturates at around 2.5 meters. This likely occurs to conform to road lane width standards of road vehicles. Increasing $L_y$ improves the stability against toppling, but we also observe that it is often correlated with an increase in height, thereby countering that stabilising advantage. The largest dimension of potential UFDs ($L_x$) spans over several orders of magnitude (Figure 6A), from e-scooters and small household furniture to large heavy vehicles. The clustering of items according to their group indicates the consistency of the definition proposed in Table 2. The characterised UFDs raw data is available online (Bayón et al., 2024).

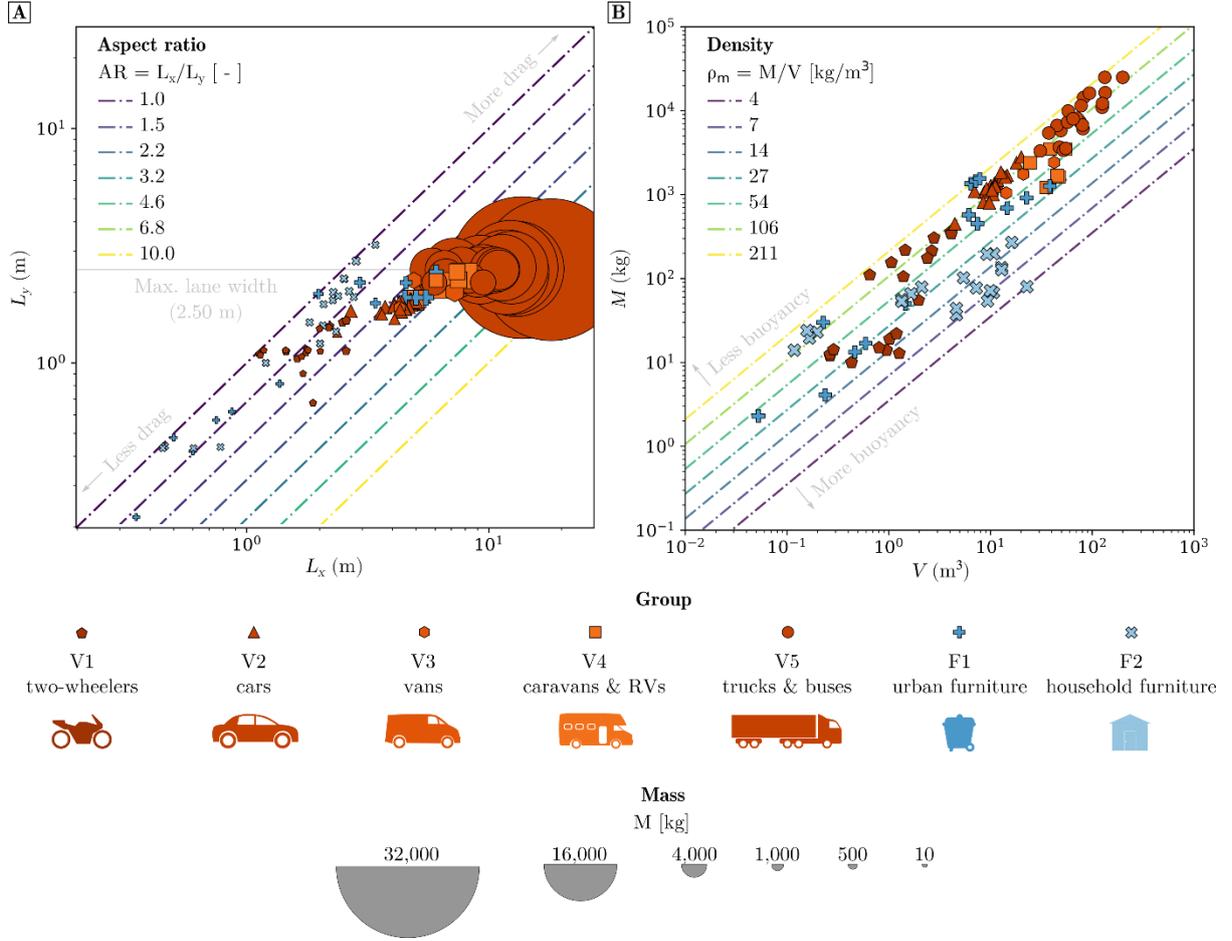

*Figure 6.* A) Distribution of the main geometrical properties of UFDs and their mass, and B) densimetric distribution of UFDs.

Figure 6B shows the dispersion of mass and volume for the potential UFDs considered, with densities covering four orders of magnitude. However, to a certain extent, we can observe clustering related to the UFD groups identified in Table 2. For instance, vehicles with motors (V2-V5) roughly remain within bulk densities of 50 to 200 kg/m$^3$, whereas urban (F1) and household furniture (F2) are considerably lighter –in terms of bulk density, 4 to 60 kg/m$^3$–, which indicates that they are considerably more susceptible to flotation instability. Urban furniture, such as rubble containers or dumpsters, is denser than household furniture, such as garden sheds or fluid tanks. While the latter may be expected to be more unstable under flood conditions, they can also be expectedly placed within less exposed areas. The mass of potential UFDs spans across nearly five orders of magnitude, whereas cars, the most studied group in scientific literature, only have masses commonly ranging from 1,100 to 1,600 kg (inter-quartile range). From a hazard-related perspective, it is interesting to note that heavy vehicles (V5) and caravans (V4) have bulk densities comparable to cars (V2), despite being heavier in mass-terms. This implies that, after the water level passes their clearance (20 to 35 cm), these vehicles can be as unstable to flotation instability as cars.

The data compiled in Figure 6 is the basis for the following stability analysis of the UFDs prevalent in the investigated floods. Based on the forces of Equation 1, we define two dimensionless ratios to study stability of potential UFDs for $h \in [0, L_z]$:

$$\frac{F_b}{F_g} = \phi_1 \frac{\rho_w}{\rho_m} \frac{h}{L_z} \qquad (2)$$

$$\frac{F_d}{F_g} = \phi_2 \frac{\rho_w}{\rho_m} \frac{C_D}{2g} \frac{hv^2}{L_y L_z} \qquad (3)$$

The first ratio represents the relationship between buoyancy (destabilising) Archimedes force and gravity (stabilising) force (weight), whereas the second represents the ratio between the force due to the flow drag (destabilising) force and the gravity force. Hence, as the values for these two ratios increase, potential UFDs become less stable due to (1) buoyancy and (2) drag forces, respectively.

Taking as reference the critical flow conditions that destabilise cars (V2), for instance from the 1:1 scale tests of Smith et al. (2019b), we can assess through Equations 2 and 3 which other UFDs in our inventory, with key physical properties available under Bayón et al. (2024) and partly presented in Figure 6, are more prone to being transported during floods. This can be estimated by identifying larger relative buoyancy (Equation 2) and drag (Equation 3), when compared to cars at flow conditions leading to their inception of movement. This allows us to propose a stability diagram (Figure 7) that considers the mobility of UFDs, to be used in the evaluation of their flood hazard.

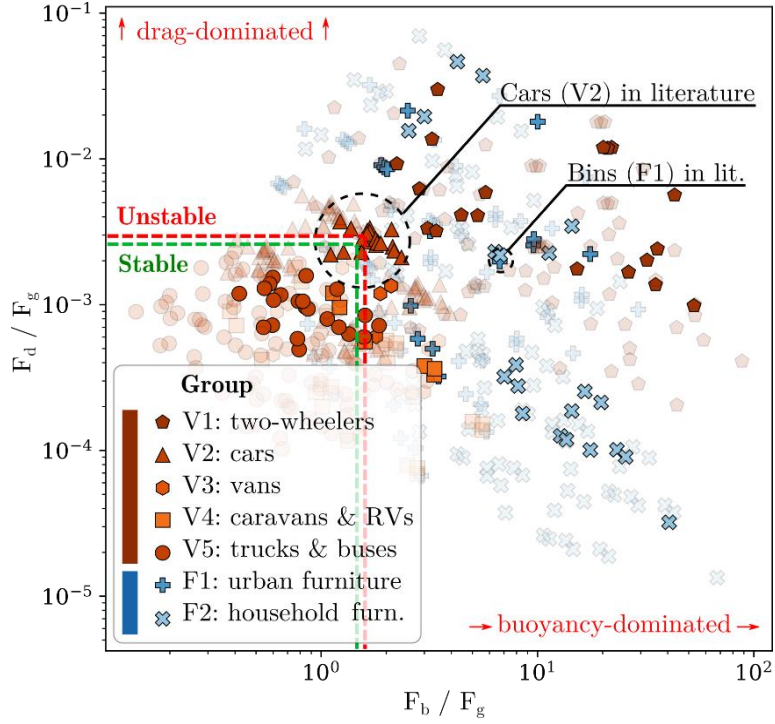

***Figure 7.*** *Simplified-stability analysis based on Equations 2 and 3. In solid, UFDs (markers following Figure 6) for flow conditions $h = 0.3$ m, $v = 2$ m/s and $\phi_1 = \phi_2 = 1$, corresponding to critical conditions for cars in Smith et al. (2019b); these are the basis for the establishment of the stable/unstable regions. On the background, other critical conditions are considered for sensitivity analysis, comprehending: $h = 0.2$ m, $v = 3$ m/s to $h = 0.5$ m, $v = 1$ m/s, with $\phi_1 = \phi_2 = 1$ to $\phi_1 = \phi_2 = 0.5$. The studied UFDs correspond to hydrodynamic stability literature (Smith et al., 2019b; Kramer et al., 2016; Xia et al., 2014; Martínez-Gomariz et al., 2020).*

Figure 7 shows the characterised UFDs under flow conditions for which cars become unstable: flow depth $h = 0.3$ m and velocity $v = 2$ m/s (based on Smith et al., 2019b, large scale tests). For simplicity, we assume $\phi_1 = \phi_2 = 1$ for all UFDs considered, corresponding to a rectangular prismatic geometrical simplification, with $C_D = 1$ as well. To assess the sensitivity of our analysis to flow conditions, we consider further flow conditions, including increased buoyancy ($h = 0.5$ m, $v = 1$ m/s) and increased flow velocity ($h = 0.2$ m, $v = 3$ m/s), for which cars became unstable in the study of Smith et al. (2019b). Additionally, to assess sensitivity to parametrisation, we consider as well $\phi_1 = 0.5$ and $\phi_2 = 0.5$ within the data shown in Figure 7.

Our analysis, illustrated in Figure 7, corroborates that only a small set of critical UFDs has been explored so far, leaving a significant portion that contributes to hazards largely unnoticed. Furthermore, Figure 7 indicates when transport triggering is buoyancy-dominated or drag-dominated, which explains if high water levels or large velocities would trigger their movement. For instance, we observe that caravans and RVs (V4) are considerably more sensitive to buoyancy than cars (V2). Heavy vehicles have a flood-mobility only slightly better than cars, whereas their clogging potential (see $L_x$ and $M$ in Figure 6A) is notably larger. Also, the majority of urban (F1) and household furniture (F2) are unconditionally more unstable than cars, either due to buoyancy (Figure 7, right side) or drag (Figure 7, upper side).

### 4. Conclusions

Large objects can be entrained and transported by urban floods, thus enhancing flood hazard where consequences are largest. In this study, we collect and analyse post-flood imagery available online to identify and classify UFDs by type according to their main characteristics, functionality and material (see Table 2). We build a dataset containing 270 images covering 63 major floods occurring in 46 different countries over the past quarter-century (1999-2023). In these, we identified 756 UFDs (data values sourced from the 270 images) which are considered in our frequency and prevalence analysis. Our UFDs classification allows clustering of certain properties and eases the study of such a heterogeneous problem.

Our analysis shows that cars, the most studied group of UFD in literature, shows up in 31.5% of all the observed images showing UFDs. Although they are the most frequent type of vehicle, larger objects such as vans, caravans, RVs, and heavier vehicles can hold higher clogging potential and are together 41% as prevalent as cars. Urban and household furniture show a non-negligible prevalence (Figure 4), whereas plastic, construction, metal debris and wood (natural or anthropogenic), among other smaller waste, are the most frequent UFDs, being present (at least one of them) in 90% of the collected images. We observe that, physically, UFDs can be widely heterogeneous, but a certain structure arises for each UFD category identified in Table 2. Increasing volume is usually accompanied by a larger mass, with a bulk density characteristic for each class of UFD (Figure 6B). In terms of stability, we also show that most of the large typified UFD (vehicles or not) are either more sensitive to drag or more prone to flotation than cars, therefore giving rise to an unexplored threat, except for a few studies (Martínez-Gomariz et al., 2020; Kramer et al., 2015, 2016).

Based on our survey, we propose a novel stability diagram (Figure 7) that can be used by urban planners to identify the most unstable urban elements and to prevent their contribution to flood hazard. In Figure 7, stability is parameterised according to the dominant destabilising forces: buoyancy or drag. When planning urban areas, this type of framework will aid the identification of which UFDs should be properly moored or avoided in flood-sensitive areas. Altogether, this research identifies and provides a first classification and characterization of the key objects in the urban landscape that might critically amplify flood hazard, pointing directions for future adjustment of the urban landscape towards more flood-resilient cities and potentially contributing to the development of modelling tools for a well-informed flood management.